\documentclass[aps,prb,showpacs,byrevtex,showpacs,amsmath,amssymb,twocolumn]{revtex4-1}
\usepackage{amssymb}
\usepackage{epsfig}
\usepackage{graphicx,subfigure}
\usepackage{color}

\definecolor{Red}{rgb}{1,0,0}

\definecolor{Blu}{rgb}{0,0,1}

\definecolor{Green}{rgb}{0,1,0}

\begin{document}

\title{Doping dependence of magnetic excitations of 1D cuprates as probed by Resonant Inelastic x-ray Scattering}
\author{Filomena Forte$^{1,2}$, Mario Cuoco$^{1,2}$, Canio Noce$^{1,2}$, and Jeroen van den Brink$^3$}
\affiliation{$^1$CNR-SPIN, I-84084 Fisciano (SA), Italy }
\affiliation{$^2$Dipartimento di Fisica ``E. R. Caianiello'',
Universit\`a di Salerno, I-84084 Fisciano (SA), Italy}
\affiliation{$^3$Institute for Theoretical Solid State Physics,
IFW-Dresden, D01171 Dresden, Germany}

\begin{abstract}
{We study the dynamical, momentum dependent two- and four-spin
response functions in doped and undoped 1D cuprates, as probed by
resonant inelastic x-ray scattering, using an exact numerical
diagonalization procedure. In the undoped $t-J$ system the
four-spin response vanishes at $\pi$, whereas the two-spin
correlator is peaked around $\pi/2$, with generally larger
spectral weight. Upon doping spectra tend to soften and broaden,
with a transfer of spectral weight towards higher energy. However,
the total spectral weight and average peak position of either
response are only weakly affected by doping up to a concentration
of 1/8. Only the two-spin response at $\pi$ changes strongly, with
a large reduction of spectral weight and enhancement of excitation
energy. At other momenta the higher-energy, generic features of
the magnetic response are robust against doping. It signals the
presence of strong short-range antiferromagnetic correlations,
even after doping mobile holes into the system. We expect this to
hold also in higher dimensions.}
\end{abstract}
\date{\today} \maketitle


\section{Introduction}
Cuprate materials have proven a fertile ground for the study of
strong electronic correlations and quantum magnetism.
Spectroscopic techniques are powerful tools to obtain information
about these properties, since they provide direct information on
the electronic and magnetic elementary excitations, which are
related to the energy spectra, the crystal structure, and so on.
Among the spectroscopic techniques used to probe magnetic
excitations in cuprates, Resonant Inelastic X-ray Scattering
(RIXS) has gained much interest  because of  the recent increases
in energy and momentum resolution, due to the enhanced brilliance
of synchrotron x-ray sources and the advances in instrumentations.
This has placed RIXS at the forefront in the study of the momentum
dependent electronic and magnetic responses over a wide energy
range.\cite{Kao1996,Hill1998,Kuiper1998,Abbamonte1999,Hasan2000,Hasan2002,Kim2002,
Kim2004,Ghiringhelli2004,Ishii2005,Ishii2005b,Grenier2005,Lu2005,Collart2006,Wakimoto2009}

In the RIXS process, x-ray radiation is inelastically scattered by
the matter and the change in energy, momentum and polarization can
be related to intrinsic excitations in the material. This process
is resonant because the energy of an incoming photon is tuned to
match an element absorption edge, thus allowing a large
enhancement of the scattered intensity. In this way the x-ray
photons can couple to charge, spin and orbital degrees of
freedom.\cite{Hill2008,Braicovich2009,Schlappa2009,Ament2009,Braichovich2010,Ellis2010,Guarise2010,Kotani2001,Ament2011,Braichovich2010b,Brink2007,Forte2008,
Haverkort2010,Igarashi2007,Igarashi2011}

In the context of the cuprates, it is by now well-established that
RIXS can detect the momentum dependence of charge excitations that
are related to the electrons and holes in the $d$
shell,\cite{Kotani2001,Ament2011,Hill1998}, but it has also been
proved, both experimentally and theoretically, that RIXS is
sensitive to the magnetic excitations of cuprates.

\subsection{Magnetic RIXS in cuprates}
In the magnetic sector, RIXS can both create single and double
spin flip excitations, corresponding to single- and bi-magnon
excitations in ordered Heisenberg antiferromagnets (AFM).
Moreover, it has been also predicted that three-magnon scatterings
contribute substantially to the magnetic spectral
weight,\cite{Ament} whereas single-magnon excitations were
observed in direct RIXS experiments at the Cu L$_3$ edge on a thin
film of La$_2$CuO$_4$,
\cite{Ament2009,Braichovich2010,Braichovich2010b,Haverkort2010}
and also on small crystals of
Sr$_2$CuO$_2$Cl$_2$.\cite{Guarise2010}

From a theoretical point of view, when the magnetic moment lies in
the plane of the $x^2-y^2$ orbital, direct spin-flip scattering in
cuprates is allowed for symmetry reasons. Thus, at least for this
class of materials, L-edge RIXS can be placed on the same footing
as neutron scattering, and both are related to the two-spin
dynamical correlation function.\cite{Ament2009}

At the Cu K-edge, the RIXS process is indirect and single-magnon
scattering is forbidden. In this case, magnetic excitations turn
out to be due to bi-magnon, as observed in insulating and doped
La$_{2-x}$Sr$_x$CuO$_4$ and
Nd$_2$CuO$_4$.\cite{Hill2008,Ellis2010} Subsequently, excitations
with a bi-magnon-like dispersion have been observed in cuprates
with high-resolution L-edge
RIXS,\cite{Braicovich2009,Schlappa2009} and M-edge
RIXS.\cite{Freelon} This makes RIXS thus complementary to optical
Raman scattering, which also measures the bi-magnon, but only at
zero momentum
transfer.\cite{Salamon1995,Blumberg1996,Naeini199,Sugai2001,Machtoub2005}.

The microscopic mechanism by which the bi-magnons couple to the
intermediate state core-hole in Cu K-edge RIXS is by the core-hole
locally modifying the superexchange constant. This leads to the
measurement of a four-spin correlation function that can be
derived in detail via the Ultrashort Core-hole Lifetime (UCL)
expansion.\cite{Brink2007,Forte2008,Brink2006,Ament2007} This
approach yields a momentum dependence of the cross section in
agreement with experiments on undoped cuprates in the
N$\mathrm{\grave{e}}$el state. In particular it reproduces the
lack of intensity at $q$=(0,0) and
($\pi,\pi$).\cite{Hill2008,Ellis2010}  At the Cu L- and M-edges,
the coupling mechanism to bi-magnon excitations is similar,
resulting in the same cross section to lowest order in the
perturbing effective potential between the spin and the core-hole,
due to the fact that the superexchange in the intermediate state
is, in this case, not just altered but completely blocked
locally.\cite{Braicovich2009}

Despite the success in describing undoped cuprates, the doping
dependence of both single- and multi-magnon measured by RIXS is
rather little unexplored, however with the first data available
showing intriguing behavior. For K-edge measurements on
La$_{2-x}$Sr$_x$CuO$_4$, the intensity of the spectral features is
found to generally decrease as $x$
increases.\cite{Hill2008,Ellis2010} The persistence of the highest
peak at $(\pi,0)$, located around 500 meV, for doping values $x$ =
0.07, well into the superconducting phase, shows that this
excitation survives even if long-range magnetic order is absent,
as long as significant short-range magnetic correlations are
present, which is well-known to be the case in the superconducting
state of LSCO.\cite{Wakimoto2004,Lipscombe2007} The evolution upon
doping observed in L-edge RIXS measurements on LSCO is even more
captivating, showing the existence of a high energy ``undoped"
branch in addition to the lower-energy dispersive features
measured in neutron scattering, which can be related to the
presence of a stripe liquid\cite{Braichovich2010b}.

\subsection{Aim and Outline}
In this paper we analyze the doping dependence of magnetic RIXS in
cuprates, and particularly we explore the evolution of the
magnetic excitations with doping. We focus on 1D systems because
i) strong quantum fluctuations are present, that maximally effect
the magnetic ordering, ii) the existence of an exact theoretical
results for magnetic excitations in the undoped 1D Heisenberg AFM,
formulated in the spinon language, provides a stringent reference
test for our numerical spectra, iii) the sampling of the Brillouin
Zone (BZ) is much more dense than in the 2D case, and iv) the
results are directly relevant for RIXS on 1D cuprates, such as
Sr$_2$CuO$_3$, where experiments probing the low-energy magnetic
excitations are entirely feasible.\cite {Seo2006} These 1D spin
liquid states may be relevant to the proposed stripe liquid
behavior in high-temperature superconductors.

We present numerical calculations of different-time Two-Spin (TS)
and Four-Spin (FS) correlation functions, evaluated on 22 sites
chain for the Heisenberg model and 16 sites chain for the $t-J$
model, in the low doping regime with a doping concentration up to
1/8.
Even if the analysis is limited by the finite length of the
system, one can extract relevant information about the intensity
and the dispersion of the magnetic features by quantitative
analysis of total spectral weight ($W_0$), and its first moment
($W_1$), corresponding to the average energy peak position. Besides,
the higher energy features are expected to be less
affected by finite size effects.

We firstly consider the undoped case assuming a description based on
the nearest-neighbour AFM Heisenberg model on a chain. It turns
out that both TS and FS correlation functions detect two-spinon
excitations, with a spectral weight that is concentrated at the
lower boundary of the continuum of excitations. We point out that
TS and FS access excitations belonging to orthogonal subspaces,
having total spin S=1 and S=0, respectively. Moreover, we recover
crucial differences about momentum dispersion and selection rules:
TS is peaked at $\pi$ where FS vanishes, whereas the latter is peaked
at $\pi/2$. This result highlights the different length-scale between the TS and the FS excitation. The latter has a characteristic length-scale of $2a$, where $a$ is the lattice spacing, as two
(neighboring) exchange bonds are broken in the intermediate state.
This shifts the maximum from $\pi$ towards  $\pm \pi/2$. These results are consistent with recent Bethe Ansatz
calculations~\cite{Klauser}. We subsequently study the spectral
evolution upon doping, by using the same approach on a $t-J$ chain
containing mobile holes. The magnetic response is found to soften
and broaden as a function of the doping concentration.
Nevertheless, both the momentum dispersion and the total spectral weight
are only slightly affected by a doping of 1/16 and 1/8. Only the
two-spin response at $\pi$ changes strongly, with a large
reduction of its spectral weight and an enhancement of the
excitation energy. At other momenta the higher-energy, generic
features of the magnetic response are robust against doping. This
shows that the strong short-range antiferromagnetic correlations
that are still present after doping give rise to higher energy,
damped magnetic excitations with considerable spectral weight.

\section{Magnetic excitations in the 1D $t-J$ model}

The $t-J$ Hamiltonian is one of the most studied model Hamiltonians
in the context of high temperature superconductivity in doped,
quasi-2D cuprates. Naturally, it is directly relevant to quasi-1D
cuprates as well, among which edge-sharing 1D spin chains in
Li$_2$CuO$_2$ and GeCuO$_3$ and corner sharing in SrCuO$_2$ and
Sr$_2$CuO$_3$. Charge excitations  have been extensively studied
by exact diagonalization of Hubbard or charge-transfer models
relevant for these systems, also in the context of
RIXS~\cite{Qian2005,Tsutsui2003}. In a Hubbard-type model for an
undoped chain the presence of singlet excitations of spinon pairs
was identified, which is due to the presence of doubly occupied
sites~\cite{Tsutsui2000}. Very recently the RIXS response of the
undoped spin-only chain was computed by Bethe Ansatz, which will
serve as a benchmark for the present exact diagonalization
study.\cite{Klauser}. However, the magnetic response of doped $t-J$
chains has received remarkably little attention in the literature
so far. Studies of spin dynamics in the 1D $t-J$ model have
considered the limit of quarter-filling, with a hole concentration
of 3/4, by Monte-Carlo\cite{assaad,deisz},
exact-diagonalization\cite{tohyama} and recursion
methods\cite{zhang}, focusing on the regime of weak and strong
coupling, comparing one- and two-dimensional features, as well as
determining the evolution of the spinon spectrum in presence of
magnetic anisotropy. Here we consider the weakly doped chain, with
doping concentrations where in the 2D cuprates superconductivity
appears and we concentrate on the specific question how far the
different types of magnetic excitations to which RIXS is sensitive
are affected by the presence of mobile charge carriers.

\subsection{Zero doping Heisenberg limit}
Without doping the 1D $t-J$ model reduces to the Heisenberg AFM,
which is one of the few many-body problems where the ground state,
which is a SU(2) singlet, and the lowest excited states are known
exactly.\cite{Bethe1931} It is described by the Heisenberg
Hamiltonian
\begin{equation}
H_{Heis}=J \sum_{\langle i,j \rangle} {\bf S}_i \cdot {\bf S}_j\ ,
\end{equation}
where $J$ is the exchange coupling and ${\bf S}_i$ is the spin
operator at the $i$-site. It is well-known that this model fails
to develop long-range N$\mathrm{\grave{e}}$el ordering (where
neighbouring spins point anti-parallel to each other) even at the
lowest temperatures but rather it has an algebraically decaying
spin-spin correlation.\cite{Luther1975,Lin1991} Moreover, the
basic excitations are
spinons,\cite{Faddeev1981,Fowler1978,Anderson1987} which are
topological excitations that can be visualized as twists of $\pi$
in the spin order. Spinons are fractional particles that possess
spin values of S = 1/2, whose dispersion relation is given by
$e(p) =\pi/2 |\sin p|$, $p \in [-\pi, 0]$ in unit of $J$. Because
of quantum mechanics constraints, only an even number of spinons
can be created.

The lowest energy excitations are made of two spinon states, that
live within a continuum in $(k,\omega)$ defined by the kinematic
constraints of momentum and energy conservation: $k = -p_1-p_2$
and $\omega= e(p_1) + e(p_2)$. Hence, for a fixed external
momentum, there exists an interval in frequency given by the
conditions:
\begin{equation}
\omega \geq \omega_{2,L}(k)=\pi/2 |\sin k|; \omega \leq
\omega_{2,U}(k)=\pi |\sin k/2|\ ,
\end{equation}
with $k \in [0, 2\pi]$. Here, $\omega_{2,L}(k)$ and
$\omega_{2,U}(k)$ correspond to the lower and upper bands,
respectively.
This behavior obviously very different from
2D and 3D systems, where the excitations in the
ordered state are spin-waves, which possess a
spin value equal to one and follow a well-defined trajectory in
energy and wave-vector space in complete contrast to the
multi-spinon continuum.

Two-spinon excitations are routinely measured by inelastic neutron
scattering\cite{Kenzelmann2002,Stone2003,Zaliznyak2004,Lake2005},
but the same response function can also be measured with RIXS.
A wealth of theoretical work has been done on the TS
correlation function related to neutron structure factor and
the exact theoretical spectrum perfectly matches with experimental
results.\cite{Kohno2007,Walters2009,Thielemann2009}

\begin{figure}
\begin{center}
\includegraphics [width =8cm]{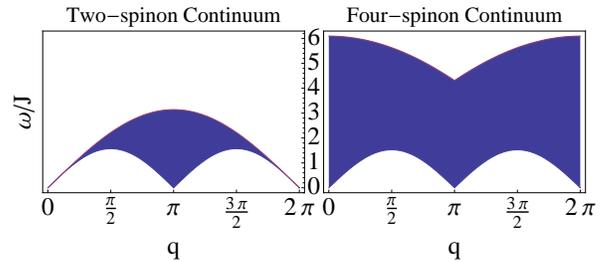}
\caption{Two-spinon (left panel) and four-spinon (right panel)
continua, stemmed in the colored regions. Color scale is not related to the spectral weight.}\label{fig:spinon-continua}
\end{center}
\end{figure}

The two-spinon part of the TS function is
finite inside the two-spinon continuum, and by construction
it vanishes identically outside of it. It has been demonstrated
that approximately two thirds of the excitations in this band are
indeed of the two-spinon type. The remaining part has a rather
small spectral weight and is carried mostly by four-spinon
excitations.\cite{Caux2006} In Fig.\ref{fig:spinon-continua},
these two-spinon and four-spinon bands are schematically represented.
Recently it was proven for the FS response function, even if it involves the
excitation of two spins, overwhelmingly fractionalizes into two-spinon states.\cite{Klauser}

Within this context, we now start a systematic
comparison of the numerical spectra related to the TS and FS
correlation functions. We will demonstrate that our results
recover the main features (e.g., dispersion and selection rules) of the
exact theoretical results, providing the starting point for the
subsequent discussion about doping evolution.

\subsection{Cross section for magnetic RIXS}
In the Introduction we mentioned that both direct and indirect
RIXS can probe magnetic excitations. In direct RIXS, single
spin-flip excitations can be made at the $2p \rightarrow 3d$ edges
of Cu because of the large spin-orbit coupling of the 2$p$
core-hole.\cite{Ament2009} Since in the intermediate state spin
and orbital angular momentum separately are no longer good quantum
numbers, orbital and spin can be exchanged and direct spin-flip
processes can in principle be allowed. As in the neutron
scattering, the RIXS cross section consists of a local structure
factor (depending on the polarization, the experimental geometry
and the excitation mechanism) multiplied
by the appropriate spin susceptibility \cite{Ament2009}\\
\begin{equation}
S(q,w) \propto \sum_f | \langle 0|S(q)|f \rangle^2 \delta(\omega
-\omega_{fi})\ , \label{Eq:correlatorS}
\end{equation}
where $|0 \rangle$ is the ground state, $| f \rangle$ an excited
state, $\omega_{fi}$ the energy lost by the photon, and
$S(q)=\sum_{i} \exp^{i q R_i} \mathbf S_{z i}$ is the single-spin
form factor. It was shown both
theoretically,\cite{Ament2009,Haverkort2010} and experimentally
\cite{Braichovich2010b} that spin-flip excitations are a result of
the effect of L$_z$S$_z$ operator in the core hole spin-orbit
coupling, so one can get a pure spin-flip transition for a
Cu$^{2+}$ ion with a hole in the $3d_{x^2-y^2}$ orbital only if
the spin is not parallel to the z-axis.

\begin{figure}
\begin{center}
\includegraphics [width =7cm]{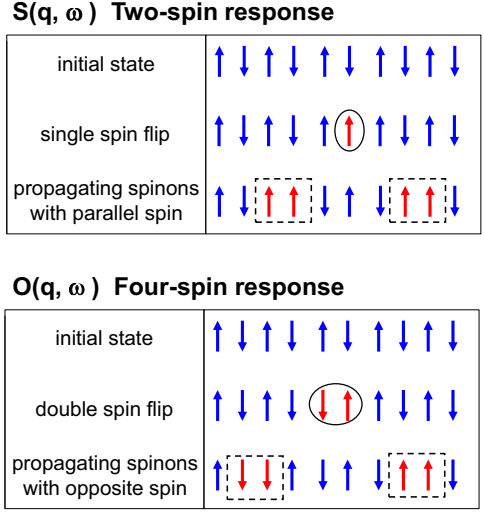}%
\caption{A schematic picture for a single spin-flip
[$S(q,\omega$)] and a double spin-flip [$O(q,\omega$)] in an AFM
Heisenberg chain. The first fractionalizes into two-spinon having
parallel spins. The latter into two-spinons carrying opposite spin
(total spin S=0). Spinons are emphasized by  the dashed box. The
circle indicates the sites where the spin flip process occurs.
}\label{Fig:two-spinon-sketch}
\end{center}
\end{figure}

For indirect RIXS at Cu K-edges ($1s \rightarrow 4p$), the core
hole couples to the spin degree of freedom locally modifying the
superexchange interactions.\cite{Brink2007,Forte2008} In this
process, the total spin of the valence electrons is conserved, and
only excitations with at least two spins flipped (with total
S$_z=0$) are allowed.\cite{Hill2008,Ellis2010} Detailed calculations of the magnetic
response functions within the UCL
expansion\cite{Brink2006,Ament2007} demonstrated that the magnetic
correlation function, measured by indirect RIXS, is a four-spin
correlation one,
\begin{equation}
O(q,w) \propto \sum_f | \langle 0|O(q)|f \rangle^2 \delta(\omega
-\omega_{fi})\ , \label{Eq:correlatorO}
\end{equation}
where  $O(q)=\sum_{i} \exp^{i q R_i} (\sum_{\delta} {\mathbf
S}_i\cdot {\mathbf S}_{i+\delta})$ is the two-spin form factor. At
the transition metal L-edges, a similar mechanism occurs (in
addition to the direct single spin-flip scattering discussed
above) because the photo-excited electron in the $3d$ subshell
frustrates the local superexchange bonds.\cite{Braicovich2009}

\begin{figure}
\begin{center}
\includegraphics [width =.9\columnwidth]{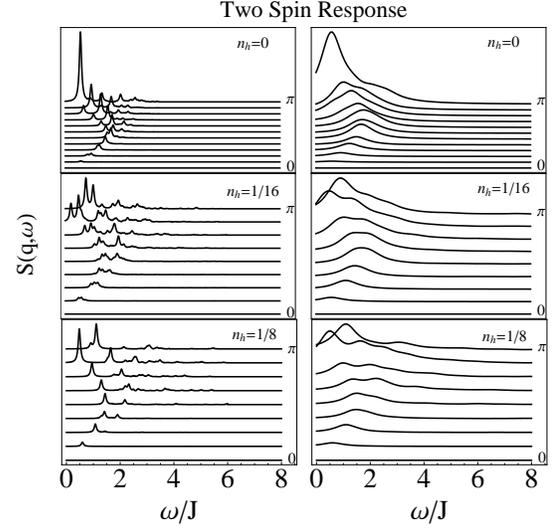}%
\caption{Inelastic part of $S(q,\omega)$ at different hole concentration associated to the
unequal time correlator defined in Eq. \ref{Eq:correlatorS},
evaluated for a 22-sites Heisenberg chain and 16-sites $t-J$
chain. For the $t-J$ model, the hopping is fixed at $t=3 J $.The
broadening of the lorentzian is $\gamma=0.5 J (0.05 J)$ for the
right (left) panel, respectively.}\label{fig:TJ-S}
\end{center}
\end{figure}
\begin{figure}
\begin{center}
\includegraphics [width =.9\columnwidth]{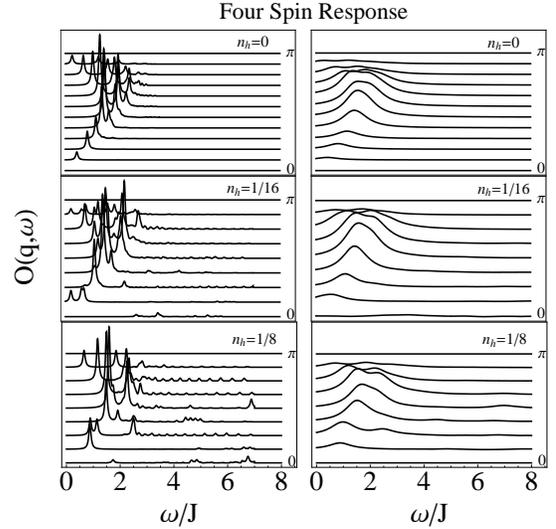}%
\caption{Inelastic part of $O(q,\omega)$ at different hole concentration associated to the
unequal time correlator defined in Eq. \ref{Eq:correlatorO},
evaluated for a 22-sites Heisenberg chain and 16-sites $t-J$ chains.
For the $t-J$ model, the hopping is fixed at $t=3 J $. The
broadening of the lorentzian is $\gamma=0.5 J (0.05 J)$ for the
right (left) panel, respectively. }\label{fig:TJ-O}
\end{center}
\end{figure}

To calculate $S(q,\omega)$ and $O(q,\omega)$, we apply the Lanczos
algorithm to a 22-sites Heisenberg chain and we employ
eigenvectors and eigenvalues for the evaluation of the correlation
functions as in Eqs. \ref{Eq:correlatorS}-\ref{Eq:correlatorO}. In
addition, we perform an extensive analysis of the momentum
dependence of the frequency moments. We evaluate the average peak
position with respect to the double differential cross section
$\frac{d^2\sigma^{(1)}}{d\Omega d\omega}$ as $W_1 \propto  \int
\omega \frac{d^2\sigma^{(1)}}{d\Omega d\omega} d\omega$ and we use
the total weight $W_0 \propto \int \frac{d^2\sigma^{(1)}}{d\Omega
d\omega}d\omega $ to gain information on the relative ratio as a
function of doping concentration and to compare the total
intensity of the bare TS and FS spectra. Note that these
outcomes are possibly easier to compare with experiments, with
respect to the differential cross-section itself. The reason is
twofold: i) the presence of statistical errors in experiments is
less relevant for the integrated quantities, so a direct comparison
with theory is feasible; ii) since lineshapes of the theoretical
spectra are typically not Lorentzian, the average excitation
energy of the calculated spectra is expected to be the most
representative theoretical result.

Before describing the results obtained within our simulation, we
would like to point out that the magnetic excitations described by
Eqs. \ref{Eq:correlatorS}-\ref{Eq:correlatorO} belong to
orthogonal subspaces, as illustrated in Fig.\ref{Fig:two-spinon-sketch}.
The figure aims to clarify that the $S(q,\omega)$ TS response is related to a single spin-flip that gives rise to a
two spinons carrying parallel spin. Instead $O(q,\omega)$ FS response function is due to a double spin-flip that
fractionalize into two spinons, having opposite spin.
Starting from the SU(2) singlet ground
state, the excitation governing $S(q,\omega)$ thus carries S=1 while
for $O(q,\omega)$ the excitation has S=0. This crucial aspect
allows to distinguish between `polarized' and `unpolarized'
excitations and suggests that the two magnetic responses can have
a different sensitivity to for example spinon-spinon interactions or an external magnetic field.

In the top panels of Figs.~\ref{fig:TJ-S}-\ref{fig:TJ-O}, we report
the undoped inelastic intensity evaluated numerically starting
from TS and FS correlation functions, respectively. There are
several common features emerging: the dispersions show
similar behavior, the excitation spectrum is mainly located in
the two-spinon band, both for $S(q,\omega)$ and $O(q,\omega)$,
with an energy scale $\omega_{2,L}<\omega<\omega_{2,U}$. Moreover,
the intensity is dominated by the spectral weight at the lower threshold of the two-spinon continuum.
These results confirm that both $S(q,\omega)$ and
$O(q,\omega)$ mainly fractionalize in two-spinon excitations\cite{Klauser}. From
a closer inspection at the energy region comprised between the
upper boundary of the two-spinon continuum and the upper boundary
of the four-spinon continuum, high energy tails emerge as
consequence of the four-spinon part of the structure factors, that
are finite but very small and rapidly approaching to zero.

\begin{figure}
\begin{center}
\includegraphics [width =1.05\columnwidth]{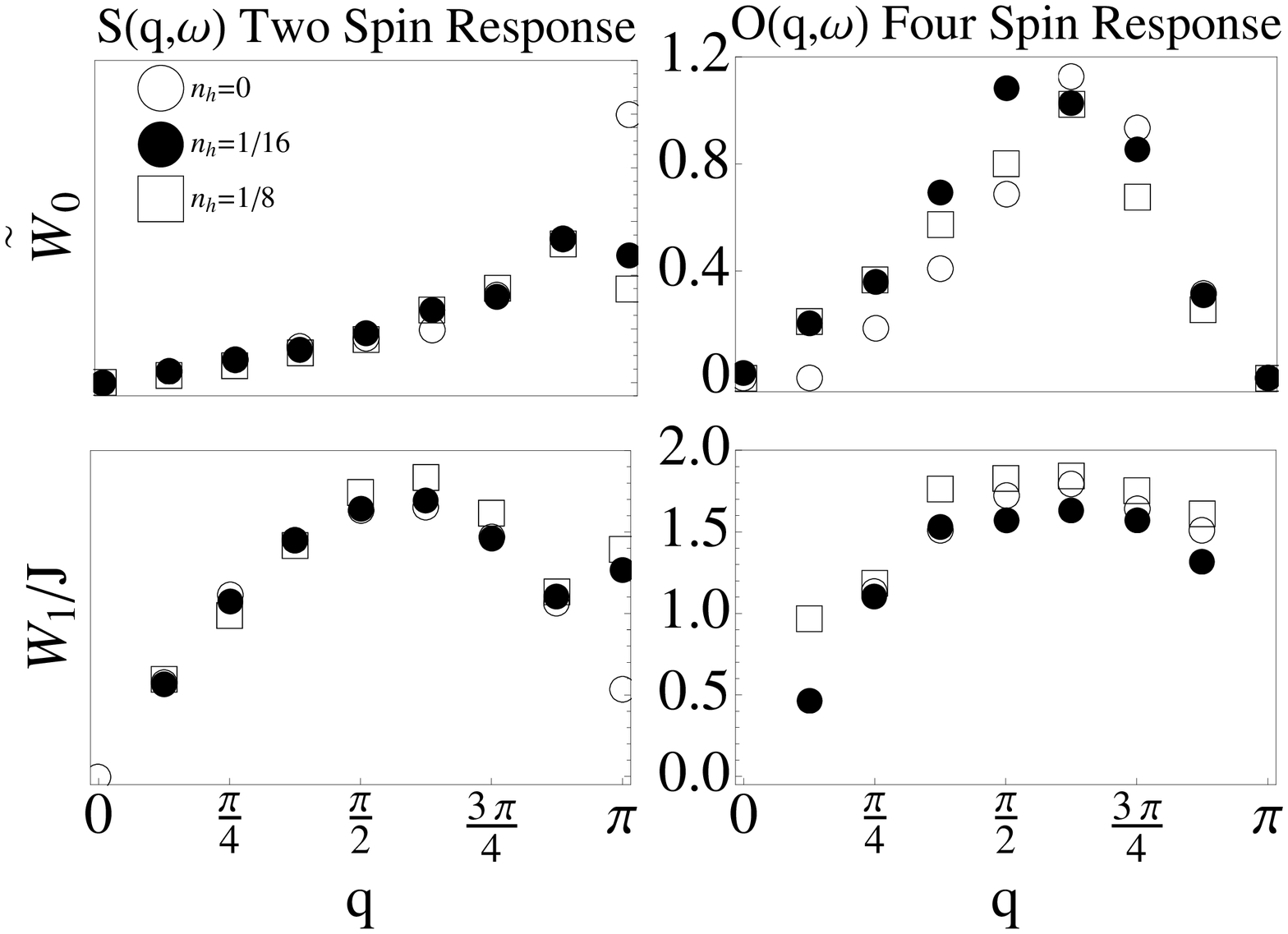}%
\caption{Total weight (upper panel) and first moment (lower panel)
of the TS (left side) and FS (right side) correlation function on
a 16-sites chain, for different doping concentrations. For
convenience of graphical representation, $W_0$ is scaled to the
total weight at $\pi$. Missing points for $W_1$ are due to
vanishing spectral weights.}\label{fig:1D-Heis-Moments}
\end{center}
\end{figure}

Nevertheless, spectra do show differences in dispersion. Looking
at the white circles of Fig. \ref{fig:1D-Heis-Moments}, we infer
that for the correlation function $O(q,\omega)$  the spectrum
disperses from zero at $q=0$, up to a maximum of $\sim 2J$ at
$\pi/2$, the intensity is suppressed at $\pi$ and nearby, a fact
that is due to structure factor $e^{iqR_i}$ and not to the density
of states, as reported in lower panel of Fig.
\ref{fig:1D-Heis-Moments} where the moment $W_1$ is plotted. The
vanishing of the RIXS intensity described by $O(q,\omega)$ at the
antiferromgnetic wavevector $\pi$ is due to the cancellation of
the sums over the two sublattices in an antiferromagnetic order. A
more detailed explanation can be found in Ref. \cite{Forte2008}.
On the contrary, $S(q,\omega)$ has its maximum at $\pi$ (see Fig.
\ref{fig:1D-Heis-Moments}). The appearance of a satellite peak at
the AFM point located at higher energies can also be noticed.
Finally, we remark that, as long as the bare correlation functions
are considered, the total weight is larger in the FS case, almost
all over the BZ (away from $\pi$), while the highest peak
intensities are comparable.

\subsection{Doping dependence}
The doped Heisenberg model is here considered in terms of the
single-band $t-J$ model,\cite{Zhang1988}
\begin{eqnarray}
H_{t-J}=&-&t\sum_{\langle i,\delta\rangle, \sigma}
\biggl(\widetilde{d}^{\dagger}_{i, \sigma}
\widetilde{d}_{i+\delta,
\sigma}+h.c.\biggr)\nonumber\\
&+&J \sum_{\langle i,j \rangle} {\bf S}_i \cdot {\bf S}_j\ ,
\end{eqnarray}
where the sums run over all the $\langle i,j \rangle$ bonds,
counted once, and $\widetilde{d}$ operators describe electrons in
the $d$ electronic levels with the constraint of no double
occupancy.

By following the same procedure described in the previous section,
we analyze the evolution upon doping of the TS and FS spectra. Two
cases are considered: the hole concentration $n_h$=1/16 and
$n_h$=1/8, respectively. In Figs. \ref{fig:TJ-S} and
\ref{fig:TJ-O} we report the spectra for
$t/J=3$, a ratio that is typical for corner-sharing cuprate perovskites.
 As one can see, the spectra are generally softened and
broadened and the intensity of the highest peaks is reduced as
n$_h$ is increased. This effect is more dramatic at $\pi$ for TS,
where the peak is strongly damped. In contrast, the doping does
not affect substantially the dispersion of the excitation
continuum, the only effect being a transfer of spectral weight
towards the upper threshold of the two-spinon continuum, with
tails in the four-spinon region. Concerning the selection rules,
the inelastic intensity is vanishing at $q=0$ for both TS and FS,
and it is suppressed at $\pi$ for $O(q, \omega)$.

Considering the frequency moments reported in Fig.
\ref{fig:1D-Heis-Moments}, interesting features come out. Namely,
as long as the TS is concerned, major differences occur around the
AFM wave-vector, where $W_0$ strongly decreases with the increases
of the doping. This result is somewhat expected since AFM
correlations are weakened by doping. The evolution of $W_1$
suggests that the average peak position stays almost unchanged
upon doping; nevertheless it is shifted to $\omega \sim 2J$
approaching $\pi$. As compared to the TS correlation function, the
intensity of the FS appears to be renormalized in a less dramatic
way. The total weight is even increased in the first half of the
BZ (see Fig. \ref{fig:1D-Heis-Moments}) and the intensity of the
highest peak at $\pi/2$ is much less softened than the
corresponding TS at $\pi$. A further look at $W_1$ allows to
conclude that $n_h$=1/16 leaves the dispersion of the average peak
position unmodified while, by increasing $n_h$, the dispersion is
shifted to slightly higher energies ($\omega \sim 2.5 J$). From
these outcomes, we can conclude that both the magnetic dispersions
mapped by $S(q,\omega)$ and $O(q,\omega)$ are quite robust against
the doping, as long as the lightly doped regime is considered, and
we also infer that the robustness of the two-spinon feature is
directly linked to the existence of short-range AFM correlations,
irrespective of doping concentration.

Finally, concerning the transfer of spectral weight towards the
upper threshold of the two-spinon continuum, we may deduce that it
can be related to an `itinerancy effect' due to the fact that the
introduction of mobile carriers induces charge fluctuations that
couple to spin excitations. We note that an analogue effect is
known to be played by correlation energy. For a single-band 1D Hubbard model at half filling, it has been showed that for large values of the on-site repulsion $U$, the spin correlations are dominated by virtual hopping processes of electrons and are described in terms of a spin-1/2 Heisenberg chain; in this limit, the spectral weight of spin structure factor $S(q,\omega)$ is concentrated at the lower spinon boundary.\cite{Bhaseen2005}
If $U$ is decreased, the real hopping of electrons becomes important and eventually dominates the spin response. The electron itinerancy may influence the magnetic correlations, and it is found that the spin structure factor of the Hubbard model differs from the spectrum of the Heisenberg model since a significant spectral weight is concentrated on the upper spinon boundary. \cite{Bhaseen2005}
We therefore infer that in our case those `itinerancy corrections' arising from charge excitations may be driven by doping with holes, both on TS and FS functions.

\section{Conclusions}
For finite Heisenberg and $t-J$ chains we have calculated the two-spin (TS) and four-spin (FS) correlation
functions, which determine the magnetic RIXS spectra of cuprates. The calculation is
performed by means of the Lanczos algorithm applied to 22-sites
Heisenberg and 16-sites $t-J$ models. We firstly show that both
functions measure multi-spinon excitations, and we find that the
dominant contribution arises from the two-spinon continuum.
Depending on which function is considered, one can access to
`polarized' ($S=1$) or `unpolarized' ($S=0$) excitations,
resulting also in a different momentum dispersion. Then, as long
as the Heisenberg model is concerned, one observes that $O(q,\omega)$
peaks at $\pi/2$ while $S(q,\omega)$ peaks at $\pi$, where
$O(q,\omega)$ vanishes; away from $\pi$, the first has a total
weight lower  than the latter. When doping is introduced, both the
TS and FS spectra are softened and spread out.
The maximum peak intensities are strongly renormalized,
but the total integrated intensity is slightly modified: the spectra tend to broaden. The TS and FS spectra
show a general transfer of spectral weight towards the higher energy sectors
of the two-spinon band. The main features of  the TS and FS spectra, however,
exemplified by for instance their zeroth and first moment, show a remarkable
robustness against to doping.

\begin{acknowledgments}
The research leading to these results has received funding from
the FP7/2007-2013 under grant agreement N. 264098 - MAMA. We thank
A. Klauser and J.-S. Caux for fruitful discussions.
\end{acknowledgments}

\end{document}